# Fiber-optic temperature sensing probe using low-coherence light source

**Kanon Toda,[1] Kenta Otsubo,[1] Kohei Noda,[1,2,3] Heeyoung Lee,[4] Kentaro Nakamura,[2] and Yosuke Mizuno[1]**

[1] *Faculty of Engineering, Yokohama National University, 79-5 Tokiwadai, Hodogaya-ku, Yokohama 240-8501, Japan*

[2] *Institute of Innovative Research, Tokyo Institute of Technology, 4259, Nagatsuta-cho, Midori-ku, Yokohama 226-8503, Japan*

[3] *Graduate School of Engineering, The University of Tokyo, 7-3-1 Hongo, Bunkyo-ku, Tokyo 113-8656, Japan*

[4] *Graduate School of Engineering and Science, Shibaura Institute of Technology, 3-7-5 Toyosu, Koto-ku, Tokyo 135-8548, Japan*

*Author e-mail addresses: toda-kanon-bw@ynu.jp, mizuno-yosuke-rg@ynu.ac.jp*

**Abstract:** We present a new approach for measuring fiber tip temperature using low-coherence Brillouin optical correlation-domain reflectometry, which eliminates the need for an independent reference path and does not entail specific processing of the fiber tip.

## 1. Introduction

Optical fiber temperature sensors are becoming increasingly popular due to their long-distance capabilities, lightweight nature, immunity to electromagnetic interference, and explosion-proof properties. Among the various techniques for measuring temperature using optical fibers, we focus here on methods for measuring the temperature near the tip of the optical fiber. To achieve this, previous studies have proposed several methods, such as inscribing fiber Bragg gratings (FBGs) [1,2] or long-period gratings (LPGs) [3,4] near the fiber under test (FUT) and creating a Fabry-Perot interferometer by applying special processing (such as forming air gaps or thin films) to the FUT end face [5]. However, these techniques require some form of processing near the FUT tip, leading to cost and labor issues. No method has been reported for measuring only the temperature near the tip of the FUT using off-the-shelf optical fibers without any processing. Alternatively, it is possible to measure the temperature near the fiber tip using distributed sensing techniques that provide information on the temperature variation along the fiber length [6–10], but these techniques are also costly.

In this work, we propose a method to measure only the temperature near the tip of the FUT without any special processing using the low-coherence Brillouin optical correlation-domain reflectometry (BOCDR) technique and demonstrate its feasibility in a simple experimental setup. As a proof-of-concept, we measure the temperature changes in a 40 cm region near the tip of a 13-m-long FUT, achieving a relatively high level of accuracy with a theoretical spatial resolution of 26 cm.

## 2. Principles

When light is incident on an optical fiber, Brillouin scattering occurs, in which the scattered light undergoes a frequency shift due to the acoustic phonons present in the fiber. This Brillouin frequency shift (BFS) is known to be proportional to the amount of strain or temperature change. In recent years, techniques for measuring strain and temperature distributions using BFS have received considerable attention. While some single-end-access configurations, such as Brillouin optical time-domain reflectometry (BOTDR) [6] and Brillouin optical frequency-domain reflectometry (BOFDR) [7], have been developed, here let us focus on BOCDR [8–11]. In standard BOCDR, a narrow-band semiconductor laser is used as a light source, and its output light (reference light) is modulated in frequency by a sinusoidal wave and interfered with the Brillouin-scattered light from the FUT. This generates a correlation peak on the FUT that acts as a measurement point, and the strain or temperature at that position can be measured from the BFS. To achieve distributed measurement, the correlation peak is scanned along the FUT by sweeping the modulation frequency. A relatively long delay line is often inserted in the reference light path so that zeroth-order correlation peak that cannot be scanned by frequency modulation may not be created in the FUT. The spatial resolution $\Delta z$ is determined by the modulation frequency $f_m$ and the modulation amplitude $\Delta f$ as [8,9]

$$\Delta z = \frac{c \Delta v_B}{2\pi n f_m \Delta f} \quad (1)$$

where $c$ is the velocity of light in vacuum, $\Delta v_B$ is the Brillouin bandwidth, and $n$ is the refractive index of the fiber core. The measurement range $d_m$ is given by the interval of neighboring correlation peaks as [8]

$$d_m = \frac{c}{2n f_m} \quad (2)$$

As one of the special implementations of BOCDR, low-coherence BOCDR has been reported [12,13]. The key difference between the low-coherence BOCDR and the standard BOCDR lies in the use of a low-coherence light source, where the driving current is noise-modulated. The noise modulation enables the interference of the reference light and the Brillouin scattered light only at the point where their optical paths are equal in the FUT, where a single correlation peak (zeroth order) is generated. By varying the length of one of the optical paths (typically, the reference arm) with a variable delay line, the zeroth order correlation peak can be scanned along the FUT, allowing for distributed sensing. The spatial resolution of the low-coherence BOCDR is determined by the coherence length of the light source, which is given by

$$\Delta z = \frac{2 \ln 2}{\pi} \cdot \frac{c}{n\, w_L}, \quad (3)$$

where $w_L$ is the laser linewidth. Thus, the spatial resolution of the low-coherence BOCDR is determined by a different principle than that of the standard BOCDR.

Here, we propose a method to implement a temperature sensing probe using the low-coherence BOCDR technology. In this method, we eliminate the independent reference light path and use the Fresnel reflected light from the FUT tip as the reference light. The configuration is shown in Fig. 1. The common-path-length point between the reference light and the Brillouin scattered light is located at the FUT tip. This is expected to generate the zeroth-order correlation peak centered at the tip, making the tip region a temperature-sensitive area. By using this configuration, a temperature sensing probe that can measure only the temperature near the tip without requiring special processing of the FUT can be implemented. This probe has the general advantages of optical fiber sensors such as strong resistance to electromagnetic interference and explosion-proofness, and can be constructed simply, reducing the cost and effort of processing.

## 3. Experiments

First, we measured the linewidth of the laser noise-modulated at different voltages using an optical spectrum analyzer

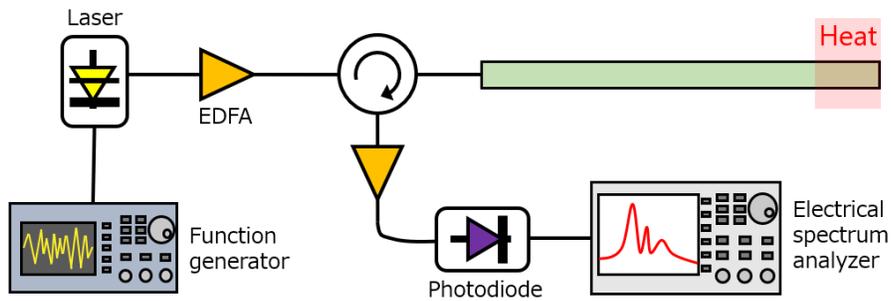

**Fig. 1.** Experimental setup of temperature sensing probe.

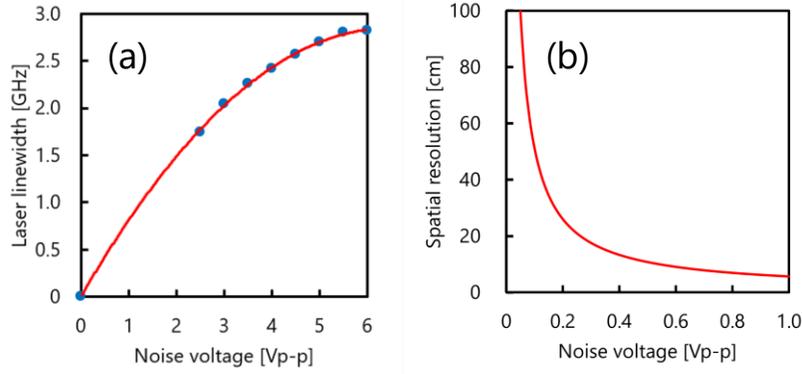

**Fig. 2**. Noise voltage dependencies of (a) the measured laser linewidth and (b) the nominal spatial resolution.

(OSA; AQ6370, Yokogawa Electric). The direct-current bias was set to 7.0 V. However, when the peak-to-peak voltage of the noise was less than 2.5 $V_{p-p}$, the linewidth of the laser did not meet the resolution of the OSA, making accurate measurement difficult. Therefore, we directly measured the laser linewidth when the noise amplitude was between 2.5 and 6.0 $V_{p-p}$, and the value at 0 $V_{p-p}$ (approximately 3 MHz) was quoted from the specification and plotted for fitting. The result is shown in Fig. 2(a), which indicates that the laser linewidth exhibits an increase as the noise voltage is augmented, albeit with a gradually decreasing dependence. Then, the spatial resolution of the temperature sensing probe was evaluated. As the correlation peak is generated at the FUT tip, the theoretical spatial resolution should be half the coherence length inside the FUT. Therefore, it can be calculated as 1/2 of Eq. (3):

$$\Delta z = \frac{ln2}{\pi} \cdot \frac{c}{n\, w_L}. \qquad (4)$$

On the basis of Eq. (4), the spatial resolution $\Delta z$ was estimated from the laser linewidth $w_L$, and the results are presented in Fig. 2(b). The figure shows that the nominal spatial resolution becomes higher (the value decreases) with increasing noise voltage, but the dependence becomes increasingly weak at higher noise voltages. This serves as fundamental data for setting the spatial resolution from the noise amplitude.

Subsequently, we demonstrated temperature measurement using the proposed probe configuration. A direct current bias of 7 V was applied to the laser, and 0.2 $V_{p-p}$ noise was added. The theoretical spatial resolution at this noise level was estimated to be approximately 26 cm according to Fig. 2(b). We used a 13-m-long silica single-mode fiber (SMF; with a BFS of ~10.8 GHz at room temperature) as the FUT with an open physical-contact (PC) connector at the tip. The 40-cm-long section of the FUT at the tip was placed in a temperature-controlled bath and heated from 25°C to 70°C, and the Brillouin gain spectra (BGSs) were observed using an electrical spectrum analyzer.

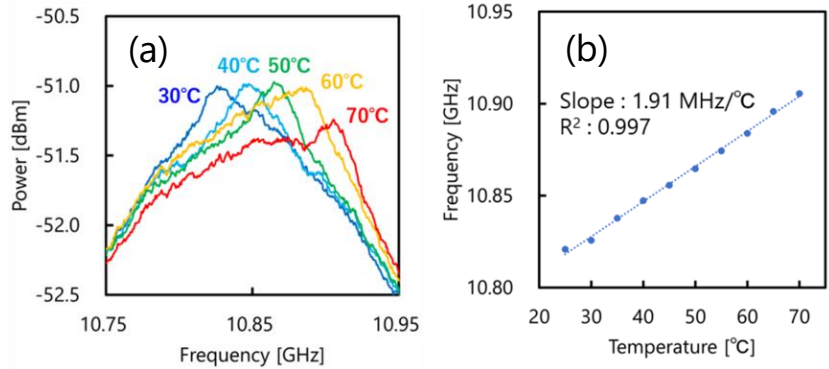

**Fig. 3**. Temperature dependencies of (a) the BGS and (b) the BFS, measured when the 0–40 cm section from the fiber tip was heated.

The observed temperature dependencies of the BGS and BFS are shown in Fig. 3(a) and (b), respectively. With an increase in the temperature of the distal 40 cm of the FUT, the component near the peak of the BGS shifted to the high-frequency side. Among the BGS components, those that did not depend on the temperature of the FUT distal end are the accumulation of the BGS generated outside the distal end of the FUT, which can be regarded as a unique noise structure. The measurable highest temperature of this probe will be determined by the condition where the power of the tip-temperature-dependent component becomes lower than the peak power of this noise structure. The BFS showed linear dependence on temperature with a coefficient of approximately 1.9 MHz/°C. The error with respect to the linear approximation was small (coefficient of determination $R^2 = 0.997$), indicating that the temperature of the distal 40 cm can be correctly measured.

## 4. Conclusions

Previous fiber-optic temperature sensing probes required some sort of processing at the tip, which presented issues with cost and labor. However, our proposed configuration utilizing low-coherence BOCDR technology allows for temperature measurement near the tip of the optical fiber with a simple experimental setup and without the need for any special processing. We demonstrated the ability to measure temperature changes applied to the 40 cm region near the tip with a relatively high level of accuracy using a theoretical spatial resolution of 26 cm. While theoretically, increasing the amplitude of noise applied to the light source should improve the spatial resolution, the ultimate limit on resolution is probably restricted by the weak scattered light power of spontaneous Brillouin scattering. Further experiments are needed to clarify this point.

**Acknowledgements**

This work was partially supported by the Japan Society for the Promotion of Science (JSPS) KAKENHI (Grant Nos. 21H04555, 22K14272, and 20J22160).